\documentstyle[preprint,aps]{revtex}
\begin{document}
\tightenlines
\title{Rare radiative $\bbox{B}$ decay to the orbitally excited
$\bbox{K_2^*(1430)}$ meson}
\author{D. Ebert$^1$,
 R. N. Faustov$^2$, V. O. Galkin$^2$ and H. Toki$^3$}
\address{$^1$Research Center for Nuclear Physics (RCNP), Osaka University,
Ibaraki, Osaka 567, Japan\\
and Institut f\"ur Physik, Humboldt--Universit\"at
zu Berlin, 10115 Berlin, Germany\\
$^2$Russian Academy of Sciences, Scientific Council for
Cybernetics, Moscow 117333, Russia\\
$^3$Research Center for Nuclear Physics (RCNP), Osaka University,
Ibaraki, Osaka 567, Japan}
\maketitle
\begin{abstract}
The exclusive rare radiative  $B$ meson decay to the orbitally
excited tensor $K_2^*(1430)$ meson is investigated in the
framework of the relativistic quark model based on the
quasipotential approach in quantum field theory. The calculated
branching ratio $BR(B\to K_2^*(1430)\gamma)=(1.7\pm 0.6)\times
10^{-5}$ as well as the ratio $BR(B\to K_2^*(1430)\gamma)/BR(B\to
K^*(892)\gamma)=0.38\pm 0.08$ is found in a good agreement with
recent experimental data from CLEO.
\end{abstract}

\bigskip\bigskip

Rare radiative decays of $B$ mesons represent an important test of the standard
model of electroweak interactions. These transitions are induced by flavour
changing neutral currents and thus they are sensitive probes of new physics
beyond the standard model.
Such decays are governed by one-loop (penguin) diagrams with
the main contribution from a virtual top quark and a $W$ boson. The
statistics of rare radiative $B$ decays considerably increased since
the first observation of the $B\to K^*\gamma$ decay in 1993 by  CLEO \cite{cleo1}.
This allowed a significantly more precise determination of exclusive and inclusive
branching ratios \cite{cleo2}. Recently the first observation of rare $B$ decay to
orbitally excited tensor strange meson
$B\to K_2^*(1430)\gamma$ has been reported by
CLEO \cite{cleo2} with a branching
ratio
\begin{equation}\label{br}
BR(B\to K_2^*(1430)\gamma)=(1.66^{+0.59}_{-0.53}\pm 0.13)\times 10^{-5},
\end{equation}
as well as the ratio of exclusive branching ratios
\begin{equation}\label{ratio}
r\equiv
\frac{BR(B\to K_2^*(1430)\gamma )}{BR(B\to K^*(892)\gamma)}
=0.39^{+0.15}_{-0.13}.
\end{equation}
These new experimental data provide a challenge to the theory.
Many theoretical approaches have been employed to predict the
exclusive $B\to K^*(892)\gamma$ decay rate (for a review see
\cite{lss} and references therein). Less attention has been payed
to rare radiative $B$ decays to excited strange mesons
\cite{a,aom,vo}. Most of these theoretical approaches
\cite{aom,vo}  rely on the heavy quark limit both for the initial
$b$ and final $s$ quark and the nonrelativistic quark model.
However, the two predictions \cite{aom,vo} for the ratio $r$ in
Eq.~(\ref{ratio}) differ by an order of magnitude, due to a
different treatment of the long distance effects and, as a
result, a different determination of corresponding Isgur-Wise
functions. Only the prediction of Ref.~\cite{vo} is consistent
with data (\ref{br}), (\ref{ratio}). Nevertheless, it is
necessary to point out that the $s$ quark in the final $K^*$
meson is not heavy enough, compared to the $\bar \Lambda$
parameter, which determines the scale of $1/m_Q$ corrections in
heavy quark effective theory \cite{n}. Thus the $1/m_s$ expansion
is not appropriate. Notwithstanding, the ideas of heavy quark
expansion can be applied to the exclusive $B\to
K^*(K_2^*)\gamma$  decays. From the kinematical analysis it
follows that the final $K^*(K_2^*)$ meson bears a large
relativistic recoil momentum $\vert {\bf\Delta} \vert$ of order
of $m_b/2$ and an energy of the same order. So it is possible to
expand the matrix element of the effective Hamiltonian both in
inverse powers of the $b$ quark mass for the initial state and in
inverse powers of the recoil momentum $\vert{\bf\Delta} \vert$
for the final state. Such an expansion has been realized by us
for the $B\to K^*(892)\gamma$ decay in the framework of the
relativistic quark model \cite{gf}.~\footnote{It is important to
note that rare radiative decays of $B$ mesons require a
completely relativistic treatment, because the recoil momentum of
the final meson is large compared to the $s$ quark mass.} The
obtained branching ratio for this decay was found in good
agreement with experimental data. Here we extend this analysis to
the decay $B\to K_2^*(1430)\gamma$.

In the standard model $B\to K^{**}\gamma$ decays are governed
by the contribution
of the electromagnetic dipole operator $O_7$ to the effective Hamiltonian
which is obtained by integrating
out the top quark and $W$ boson and using the Wilson expansion \cite{eh}:
\begin{equation}\label{o7}
O_7=\frac{e}{16\pi ^2}\bar s\sigma ^{\mu \nu
}(m_bP_R+m_sP_L)bF_{\mu \nu }, \qquad P_{R,L}=(1 \pm \gamma _5)/2.
\end{equation}
The matrix elements of this operator between the initial $B$ meson state and
the final state of the orbitally
excited tensor $K_2^*$ meson have the following covariant decomposition
\begin{eqnarray}\label{ff}
\langle K_2^*(p',\epsilon)|\bar s i k_\nu \sigma_{\mu \nu}b|B(p)\rangle
&=&i g_{+}(k^2)
\epsilon_{\mu \nu \lambda \sigma } \epsilon^{*\nu \beta}\frac{p_\beta}{M_B}
k^\lambda (p+p')^\sigma , \cr
\langle K_2^*(p',\epsilon)|\bar s i k_\nu \sigma_{\mu \nu}
\gamma_5b|B(p)\rangle & = &
g_{+}(k^2)\left(\epsilon_{\beta \gamma}^*\frac{p^\beta
p^\gamma }{M_B}(p+p')_\mu
-\epsilon_{\mu \beta}^*\frac{p^\beta }{M_B}(p^2-p'^2)   \right) \cr
& & +g_{-}(k^2)\left(\epsilon_{\beta \gamma}^*\frac{p^\beta
p^\gamma }{M_B}k_\mu
-\epsilon_{\mu \beta}^*\frac{p^\beta }{M_B}k^2\right) \cr
& &+h(k^2)((p^2-p'^2)k_\mu -(p+p')_\mu k^2)\epsilon_{\beta\gamma }^*
\frac{p^\beta p^\gamma  }{M_B^2 M_{K_2^*}},
\end{eqnarray}
where $\epsilon_{\mu\nu}$ is a polarization tensor of the final
tensor meson and $k= p -p'$ is the four momentum of the emitted
photon. The exclusive decay rate for the emission of a real
photon ($k^2=0$) is determined by the single form factor
$g_{+}(0)$ and is given by
\begin{equation}\label{drate}
\Gamma(B\to K_2^*\gamma)=
\frac{\alpha }{256\pi^4} G_F^2m_b^5|V_{tb}V_{ts}|^2
|C_7(m_b)|^2 g_{+}^2(0) \frac{M_B^2}{M_{K_2^*}^2}
\left(1-\frac{M_{K_2^*}^2}{M_B^2}
 \right)^5\left( 1+\frac{M_{K_2^*}^2}{M_B^2}\right),
\end{equation}
where $V_{ij}$ are the  Cabibbo-Kobayashi-Maskawa matrix elements and
$C_7(m_b)$ is the Wilson coefficient in front of the operator $O_7$.
It is convenient to consider the ratio of exclusive to inclusive
branching ratios, for
which we have
\begin{equation}\label{rk2}
R_{K_2^*}\equiv
\frac{BR(B\to K_2^*(1430)\gamma)}{BR(B\to X_s\gamma)}=
\frac18 g_{+}^2(0)\frac{M_B^2}{M_{K_2^*}^2}
\frac{\left(1-{M_{K_2^*}^2}/{M_B^2}
 \right)^5\left( 1+{M_{K_2^*}^2}/{M_B^2}\right)}{\left(1-{m_s^2}/{m_b^2}
 \right)^3\left( 1+{m_s^2}/{m_b^2}\right)}.
\end{equation}
The recent experimental value for the inclusive decay branching
ratio \cite{t}
$$BR(B\to X_s\gamma)=(3.15\pm 0.35\pm 0.41)\times 10^{-4}$$
is in a good agreement with theoretical calculations.

Now we use the relativistic quark model for the calculation of the form factor
$g_{+}(0)$. In our model a meson is described by the wave
function of the bound quark-antiquark state, which satisfies the
quasipotential equation \cite{3} of the Schr\"odinger type~\cite{4}:
\begin{equation}
\label{quas}
{\left(\frac{b^2(M)}{2\mu_{R}}-\frac{{\bf
p}^2}{2\mu_{R}}\right)\Psi_{M}({\bf p})} =\int\frac{d^3 q}{(2\pi)^3}
 V({\bf p,q};M)\Psi_{M}({\bf q}),
\end{equation}
where the relativistic reduced mass is
\begin{equation}
\mu_{R}=\frac{M^4-(m^2_q-m^2_Q)^2}{4M^3}.
\end{equation}
In the center-of-mass system the relative momentum squared on mass shell
reads
\begin{equation}
{b^2(M) }
=\frac{[M^2-(m_q+m_Q)^2][M^2-(m_q-m_Q)^2]}{4M^2}.
\end{equation}

The kernel
$V({\bf p,q};M)$ in Eq.~(\ref{quas}) is the quasipotential operator of
the quark-antiquark interaction. It is constructed with the help of the
off-mass-shell scattering amplitude, projected onto the positive
energy states. An important role in this construction is played
by the Lorentz-structure of the confining quark-antiquark interaction
in the meson.  In
constructing the quasipotential of the quark-antiquark interaction
we have assumed that the effective
interaction is the sum of the usual one-gluon exchange term and the mixture
of vector and scalar linear confining potentials.
The quasipotential is then defined by
\cite{mass}
\begin{eqnarray}
\label{qpot}
V({\bf p,q};M)&=&\bar{u}_q(p)\bar{u}_Q(-p){\cal V}({\bf p}, {\bf
q};M)u_q(q)u_Q(-q)\cr
&=&\bar{u}_q(p)
\bar{u}_Q(-p)\Bigg\{\frac{4}{3}\alpha_sD_{ \mu\nu}({\bf
k})\gamma_q^{\mu}\gamma_Q^{\nu}\cr
& & +V^V_{\rm conf}({\bf k})\Gamma_q^{\mu}
\Gamma_{Q;\mu}+V^S_{\rm conf}({\bf
k})\Bigg\}u_q(q)u_Q(-q),
\end{eqnarray}
where $\alpha_s$ is the QCD coupling constant, $D_{\mu\nu}$ is the
gluon propagator in the Coulomb gauge
and ${\bf k=p-q}$; $\gamma_{\mu}$ and $u(p)$ are
the Dirac matrices and spinors
\begin{equation}
\label{spinor}
u^\lambda({p})=\sqrt{\frac{\epsilon(p)+m}{2\epsilon(p)}}
\left(\begin{array}{c}
1\\ \displaystyle\frac{\mathstrut\bbox{\sigma p}}{\mathstrut\epsilon(p)+m}
\end{array}\right)
\chi^\lambda
\end{equation}
with $\epsilon(p)=\sqrt{{\bf p}^2+m^2}$.
The effective long-range vector vertex is
given by
\begin{equation}
\Gamma_{\mu}({\bf k})=\gamma_{\mu}+
\frac{i\kappa}{2m}\sigma_{\mu\nu}k^{\nu},
\end{equation}
where $\kappa$ is the Pauli interaction constant characterizing the
nonperturbative anomalous chromomagnetic moment of quarks. Vector and
scalar confining potentials in the nonrelativistic limit reduce to
\begin{equation}\label{vconf}
V^V_{\rm conf}(r)=(1-\varepsilon)(Ar+B),\qquad
V^S_{\rm conf}(r) =\varepsilon (Ar+B),
\end{equation}
reproducing
\begin{equation}
V_{\rm conf}(r)=V^S_{\rm conf}(r)+
V^V_{\rm conf}(r)=Ar+B,
\end{equation}
where $\varepsilon$ is the mixing coefficient.

The quasipotential for the heavy quarkonia,
expanded in $v^2/c^2$, can be found in Refs.~\cite{mass,pot} and for
heavy-light mesons in \cite{egf}.
All the parameters of
our model, such as quark masses, parameters of the linear confining potential,
mixing coefficient $\varepsilon$ and anomalous
chromomagnetic quark moment $\kappa$, were fixed from the analysis of
heavy quarkonia masses \cite{mass} and radiative decays \cite{gfm}.
The quark masses
$m_b=4.88$ GeV, $m_c=1.55$ GeV, $m_s=0.50$ GeV, $m_{u,d}=0.33$ GeV and
the parameters of the linear potential $A=0.18$ GeV$^2$ and $B=-0.30$ GeV
have the usual quark model values.
In Ref.~\cite{fg} we have considered the expansion of  the matrix
elements of weak heavy quark currents between pseudoscalar and vector
meson ground states up to the second order in inverse powers of
the heavy quark
masses. It has been found that the general structure of the leading,
first,
and second order $1/m_Q$ corrections in our relativistic model is in accord
with the predictions of HQET. The heavy quark symmetry and QCD impose rigid
constraints on the parameters of the long-range potential in our model.
The analysis
of the first order corrections \cite{fg} fixes the value of the
Pauli interaction
constant $\kappa=-1$. The same value of $\kappa$  was found previously
from  the fine splitting of heavy quarkonia ${}^3P_J$- states \cite{mass}.
The value of the parameter mixing
vector and scalar confining potentials $\varepsilon=-1$
was found from the analysis of the second order corrections \cite{fg}.
This value is very close to the one determined from considering radiative
decays of heavy quarkonia \cite{gfm}.

In the quasipotential approach,  the matrix element of the weak current
$J_\mu=\bar s \frac{i}{2} k^\nu \sigma_{\mu\nu}(1+\gamma^5)b$
between the states of a $B$ meson and an orbitally excited $K_2^*$ meson has
the form \cite{f}
\begin{equation}\label{mxet}
\langle K_2^* \vert J_\mu (0) \vert B\rangle
=\int \frac{d^3p\, d^3q}{(2\pi )^6} \bar \Psi_{K_2^*}({\bf
p})\Gamma _\mu ({\bf p},{\bf q})\Psi_B({\bf q}),\end{equation}
where $\Gamma _\mu ({\bf p},{\bf
q})$ is the two-particle vertex function and  $\Psi_{B,K_2^*}$ are the
meson wave functions projected onto the positive energy states of
quarks and boosted to the moving reference frame.
 The contributions to $\Gamma$ come from Figs.~1 and 2.
The contribution $\Gamma^{(2)}$ is the consequence of the
projection onto the positive-energy states. Note that the form
of the relativistic corrections resulting from the vertex function
$\Gamma^{(2)}$ explicitly depends on the Lorentz structure of the
$q\bar q$-interaction.  The vertex functions look like
\begin{equation}\label{gam1}
\Gamma_\mu ^{(1)}({\bf p},{\bf q})=\bar
u_s(p_1)\frac{i}{2}\sigma_{\mu \nu} k^\nu
(1+\gamma^5)u_b(q_1)(2\pi)^3\delta({\bf p}_2-{\bf q}_2),\end{equation}
and
\begin{eqnarray}\label{gam2}
\Gamma_\mu^{(2)}({\bf p},{\bf q})&=&\bar u_s(p_1)\bar
u_q(p_2)\frac{1}{ 2} \biggl\{ i\sigma_{1\mu
\nu}k_\nu(1+\gamma_1^5)\frac{\Lambda_b^{(-)}({ k}_1)}{ \epsilon
_b(k_1)+\epsilon_b(p_1)}\gamma_1^0{\cal V}({\bf p}_2-{\bf q}_2)\nonumber\\
& & +{\cal V}({\bf p}_2-{\bf q}_2)\frac{\Lambda_s^{(-)}(k_1')}{
\epsilon_s(k_1')+ \epsilon_s(q_1)}\gamma_1^0i\sigma_{1\mu
\nu}k_\nu(1+\gamma_1^5)\biggr\}u_b(q_1) u_q(q_2), \end{eqnarray}
where ${\bf k}_1={\bf p}_1-{\bf\Delta};\quad {\bf k}_1'={\bf
q}_1+{\bf\Delta};\quad {\bf\Delta}={\bf p}_{K_2^*}-{\bf p}_B$;
$$\Lambda^{(-)}(p)={\epsilon(p)-\bigl( m\gamma ^0+\gamma^0({\bf
\gamma p})\bigr) \over 2\epsilon (p)}.$$

It is important to note that the wave functions entering the weak current
matrix element (\ref{mxet}) cannot  be both in the rest frame.
In the $B$ meson rest frame, the $K_2^*$ meson is moving with the recoil
momentum ${\bf \Delta}$. The wave function
of the moving $K_2^*$ meson $\Psi_{K_2^*\,{\bf\Delta}}$ is connected
with the $K_2^*$ wave function in the rest frame
$\Psi_{K_2^*\,{\bf 0}}\equiv \Psi_{K_2^*}$ by the transformation \cite{f}
\begin{equation}
\label{wig}
\Psi_{K_2^*\,{\bf\Delta}}({\bf
p})=D_s^{1/2}(R_{L_{\bf\Delta}}^W)D_q^{1/2}(R_{L_{
\bf\Delta}}^W)\Psi_{K_2^*\,{\bf 0}}({\bf p}),
\end{equation}
where $R^W$ is the Wigner rotation, $L_{\bf\Delta}$ is the Lorentz boost
from the meson rest frame to a moving one. The wave functions of $B$ and
$K_2^*$ mesons at rest were calculated by numerical solution of the
quasipotential equation (\ref{quas}).

We substitute the vertex functions $\Gamma^{(1)}$  and $\Gamma^{(2)}$
given by Eqs.~(\ref{gam1}) and (\ref{gam2})
in the decay matrix element (\ref{mxet})  and take into account the wave
function transformation (\ref{wig}).
The resulting structure of this matrix element is
rather complicated, because it is
necessary  to integrate both over  $d^3 p$
and $d^3 q$. The $\delta$ function in expression  (\ref{gam1}) permits
us to perform one of these integrations
and thus this contribution  can be easily
calculated. The calculation  of the vertex function
$\Gamma^{(2)}$ contribution is  more difficult. Here, instead
of a $\delta$ function, we have a complicated structure, containing the
$q\bar q$ interaction potential in the meson.
However, we can expand this contribution in the inverse
powers of the heavy $b$ quark mass and large
recoil momentum $|{\bf \Delta}|\sim
m_b/2$ of the final $K^{**}$ meson.   Such an
expansion is carried out up to the second order.~\footnote{This means that
in expressions for $g_{+}^{(2)V}(0)$ and $g_{+}^{(2)S}(0)$ we neglect terms
proportional to the third order product of small binding energies and ratios
${\bf p}^2/\epsilon_s^3(\Delta)$, ${\bf p}^2/\epsilon_b^3(\Delta)$
as well as higher order terms.} Then we use the  quasipotential equation in
order to perform one of the integrations in the current matrix element.
As a result we get for the form factor $g_{+}(0)$ the following expression
with $\kappa=-1$
\begin{eqnarray}\label{gpl}
g_{+}(0)&=&g_{+}^{(1)}(0)+
(1-\varepsilon)g_{+}^{(2)V}(0)+\varepsilon g_{+}^{(2)S}(0)\\
\label{g1}
g_{+}^{(1)}(0) &=&\frac{1}{\sqrt{3}} \sqrt{\frac{E_{K^*_2}}{M_B}}
\frac{M_{K_2^*}}{E_{K^*_2}+
M_{K_2^*}} \int \frac{d^3p}{(2\pi )^3} \bar\psi_{K_2^*}
\Bigl({\bf p}+
\frac{2\epsilon_q }{E_{K^*_2}+M_{K_2^*}}{\bf \Delta } \Bigr)\cr \cr
&&\times\sqrt{\frac{\epsilon_s(p+\Delta )+m_s}{2\epsilon_s(p+\Delta )}}
\sqrt{\frac{\epsilon_b(p )+m_b}{2\epsilon_b(p )}}
\Biggl\{-3(E_{K^*_2}+M_{K_2^*})
\frac{({\bf p}\cdot {\bf\Delta})}{p{\bf \Delta}^2}\cr \cr
&&\times\left[1+\frac{M_B-E_{K^*_2}}
{\epsilon_s(p+\Delta )+m_s } \right]
+\left[\frac{p}{\epsilon_q(p)+m_q }
-\frac{p}{\epsilon_s(p+\Delta )+m_s} \right] \cr \cr
&&\times\left[1+\frac{M_B-E_{K^*_2}}
{\epsilon_s(p+\Delta )+m_s }
-\frac{{\bf p}^2}{[\epsilon_s(p+\Delta )+m_s]
[\epsilon_b(p )+m_b]} \right]\Biggr\} \psi_B({\bf p}),\\  \cr
\label{g2v}
g_{+}^{(2)V}(0) &=&\frac{1}{\sqrt{3}} \sqrt{\frac{E_{K^*_2}}{M_B}}
\frac{M_{K_2^*}}{E_{K^*_2}+
M_{K_2^*}} \int \frac{d^3p}{(2\pi )^3} \bar\psi_{K_2^*}
\Bigl({\bf p}+\frac{2\epsilon_q }{E_{K^*_2}+
M_{K_2^*}}{\bf \Delta } \Bigr)\cr \cr
&&\times\sqrt{\frac{\epsilon_s(\Delta )+m_s}{2\epsilon_s(\Delta )}}
\Biggl\{3(E_{K^*_2}+M_{K_2^*})
\frac{({\bf p}\cdot {\bf\Delta})}{p{\bf \Delta}^2}
\frac{M_B-\epsilon_b(p)-\epsilon_q(p)}{2[\epsilon_s(\Delta)+m_s]}
-\frac{p}{\epsilon_q(p)+m_q }\cr \cr
&&\times\frac{1}{2[\epsilon_s(\Delta)+m_s]^2}\biggl((M_B+M_{K_2^*})
[M_B-\epsilon_b(p)-\epsilon_q(p)]
+(E_{K_2^*}+M_{K_2^*})\cr \cr
&&\times \biggl[M_{K_2^*}-
\epsilon_s\Bigl({\bf p}+\frac{
2\epsilon_q }{E_{K^*_2}+M_{K_2^*}}{\bf \Delta }
\Bigr)-  \epsilon_s\Bigl({\bf p}+
\frac{2\epsilon_q }{E_{K^*_2}+M_{K_2^*}}{\bf
\Delta } \Bigr) \biggr]\biggr)\Biggr\} \psi_B({\bf p}),\\  \cr
\label{g2s}
g_{+}^{(2)S}(0) &=&\frac{1}{\sqrt{3}} \sqrt{\frac{E_{K^*_2}}{M_B}}
\frac{M_{K_2^*}}{E_{K^*_2}+M_{K_2^*}}
\int \frac{d^3p}{(2\pi )^3} \bar\psi_{K_2^*}
\Bigl({\bf p}+
\frac{2\epsilon_q }{E_{K^*_2}+M_{K_2^*}}{\bf \Delta } \Bigr)\cr \cr
&&\times\sqrt{\frac{\epsilon_s(\Delta )+m_s}{2\epsilon_s(\Delta )}}
\Biggl\{\Biggl[-3(E_{K^*_2}+M_{K_2^*})
\frac{({\bf p}\cdot {\bf\Delta})}{p{\bf \Delta}^2}+
\frac{p}{\epsilon_q(p)+m_q }\Biggr]\cr \cr
&&\times\Biggl(\frac{\epsilon_s(\Delta )-m_s}
{2\epsilon_s(\Delta )[\epsilon_s(\Delta )+m_s]}
 \biggl[M_{K_2^*}-
\epsilon_s\Bigl({\bf p}+\frac{
2\epsilon_q }{E_{K^*_2}+M_{K_2^*}}{\bf \Delta }
\Bigr)\cr \cr
&&-  \epsilon_s\Bigl({\bf p}+
\frac{2\epsilon_q }{E_{K^*_2}+M_{K_2^*}}{\bf
\Delta } \Bigr) \biggr]
+\frac{M_B-E_{K^*_2}}{2[\epsilon_s(p+\Delta )+m_s]^2 }
 \biggl[M_B+M_{K_2^*}-
\epsilon_b(p)-\epsilon_q(p)\cr \cr
&&-\epsilon_s\Bigl({\bf p}+\frac{
2\epsilon_q }{E_{K^*_2}+M_{K_2^*}}{\bf \Delta }
\Bigr)-  \epsilon_s\Bigl({\bf p}+
\frac{2\epsilon_q }{E_{K^*_2}+M_{K_2^*}}{\bf
\Delta } \Bigr) \biggr]\Biggr)\Biggr\} \psi_B({\bf p}),
\end{eqnarray}
where the superscripts ``(1)" and ``(2)" correspond to contributions coming
from Figs.~1 and 2,
$S$ and $V$  mean the scalar and vector potentials in Eq.~(\ref{vconf}),
$\psi_{K^*_2,B}$ are radial parts of the wave functions. The
recoil momentum and the energy of the ${K_2^*}$ meson are given by
\begin{equation}\label{delta}
\vert {\bf\Delta}\vert=\frac{M_B^2-M_{K_2^*}^2}{
2M_B};\qquad E_{K_2^*}=\frac{M_B^2+M_{K_2^*}^2}{ 2M_B}.
\end{equation}

We can check the consistency of our resulting formulas by taking
the formal limit of   $b$ and $s$ quark masses going to
infinity.~\footnote{As it was noted above  such limit is
justified only for the $b$ quark.} In this limit according to the
heavy quark effective theory \cite{vo2} the function
$\xi_F=2\sqrt{M_BM_{K_2^*}}g_{+}/(M_B+M_{K_2^*})$ should coincide
with the Isgur-Wise function $\tau$ for the semileptonic $B$
decay to the orbitally excited tensor $D$ meson, $B\to
D_2^*e\nu$. Such semileptonic decays have been considered by us
in Ref.~\cite{orb}. It is easy to verify that the equality of
$\xi_F$ and $ \tau$ is satisfied in our model if we also use the
expansion in $(w-1)/(w+1)$ ($w$ is a scalar product of
four-velocities of the initial and final mesons), which is small
for the $B\to D_2^*e\nu$ decay \cite{orb}. Using Eq.~(\ref{gpl})
to calculate the ratio of  the form factor $g_+(0) $ in the
infinitely heavy $b$ and $s$ quark limit to the same form factor
in the leading order of  expansions in inverse powers of the
heavy $b$ quark mass and large recoil momentum $|{\bf\Delta}|$
we find that it is equal to $M_B/\sqrt{M_B^2+M_{K^*_2}^2}\approx
0.965$. The corresponding ratio of form factors of the exclusive
rare radiative $B$ decay to the vector $K^*$ meson $F_1(0)$ (see
Eq.~(23) of Ref.~\cite{gf}) is equal to
$M_B/\sqrt{M_B^2+M_{K^*}^2}\approx 0.986$. Therefore we conclude
that the ratios of form factors $g_+(0)/F_1(0)$ in the leading
order of these expansions differ by factor
$\sqrt{M_B^2+M_{K^*}^2}/\sqrt{M_B^2+M_{K^*_2}^2}\approx 0.98$.
This is the consequence of the relativistic dynamics leading to
the effective expansion in inverse powers of the $s$ quark energy
$\epsilon_s(p+\Delta)=\sqrt{({\bf p+\Delta})^2+m_s^2}$, which is
large in one case due to the large $s$ quark mass and in the
other one due to the large recoil momentum ${\bf \Delta}$.  As a
result both expansions give similar final expressions in the
leading order. Thus we can expect that the ratio $r$ from
(\ref{ratio}) in our calculations should be close to the one
found in the infinitely heavy $s$ quark limit \cite{vo}.

The results of numerical calculations using formulas
(\ref{drate}), (\ref{rk2}), (\ref{gpl})--(\ref{g2s}) for
$\varepsilon=-1$ are given in Table~\ref{tb}. There we also show
our previous predictions for the $B\to K^*\gamma$ decay \cite{gf}.
Our results are confronted  with other theoretical calculations
\cite{a,aom,vo} and recent experimental data \cite{cleo2}. We
find a good agreement of our predictions for decay rates with the
experiment and estimates of Ref.~\cite{vo}. Other theoretical
calculations substantially disagree with data either for $B\to
K^*\gamma$ \cite{a} or for $B\to K^*_2\gamma$ \cite{aom} decay
rates. As a result our predictions and those of Ref.~\cite{vo}
for the ratio $r$ from (\ref{ratio}) are well consistent with
experiment, while the $r$ estimates of \cite{a}  and \cite{aom}
are several times larger than the experimental value (see
Table~\ref{tb}). As it was argued above, it is not accidental
that $r$ values in our and Ref.~\cite{vo} approaches are close.
The agreement of both predictions for branching ratios could be
explained by some specific cancellation of finite $s$ quark mass
effects and relativistic corrections which were neglected in
Ref.~\cite{vo}. Though our numerical results agree with
Ref.~\cite{vo}, we believe that our analysis is more consistent
and reliable. We do not use the ill-defined limit $m_s\to\infty$,
and our quark model consistently takes into account some
important relativistic effects, for example, the Lorentz
transformation of the wave function of the final $K^{**}$ meson
(see Eq.~(\ref{wig})). Such a transformation turns out to be very
important, especially for $B$ decays to orbitally excited mesons
\cite{orb}. The large value of the recoil momentum $|{\bf
\Delta}|\sim m_b/2$ makes relativistic effects to play a
significant role. On the other hand this fact simplifies our
analysis since it allows to make an expansion both in inverse
powers of the large $b$ quark mass and in the large recoil
momentum.

The authors express their gratitude to P. Ball, A. Golutvin and
V. Savrin for discussions. D.E. acknowledges the support provided
to him by the Ministry of Education of Japan (Monbusho) for his
work at RCNP of Osaka University. Two of us (R.N.F and V.O.G.)
were supported in part by the {\it Deutsche
Forschungsgemeinschaft} under contract EB 139/2-1 and {\it
Russian Foundation for Fundamental Research} under Grant No.\
00-02-17768.

\begin{table}
\caption{ Our results in comparison with other theoretical
predictions and experimental data for branching ratios and their
ratios $R_{K^*}\equiv\frac{BR(B\to K^*\gamma)}{BR(B\to
X_s\gamma)}$,  $R_{K_2^*}\equiv\frac{BR(B\to
K_2^*\gamma)}{BR(B\to X_s\gamma)}$, $r\equiv\frac{BR(B\to
K_2^*\gamma )}{BR(B\to K^*\gamma)}$. Our values for the $B\to
K^*\gamma$ decay are taken from Ref.~[8]. } \label{tb}
\begin{tabular}{cccccc}
Value&our&Ref.~\cite{a}&Ref.~\cite{aom}
&Ref.~\cite{vo}&Exp.  \cite{cleo2}\\
\hline
$BR(B\to K^*\gamma)\times 10^5$& $4.5\pm1.5$
& 1.35 & $1.4 - 4.9$&$4.71\pm1.79$
&$4.55^{+0.72}_{-0.68}\pm0.34^a $ \\
 & & & & & $3.76^{+0.89}_{-0.83}\pm0.28^b$  \\
$R_{K^*}$ (\%) &$15\pm3$&4.5 &$3.5 - 12.2$ &$16.8\pm6.4$ &\\
$BR(B\to K_2^*\gamma)\times 10^5$&$1.7\pm0.6$
& 1.8 & $6.9 - 14.8$
& $1.73\pm0.80$& $1.66^{+0.59}_{-0.53}\pm0.13$ \\
$R_{K_2^*}$ (\%)
& $5.7\pm1.2$ & 6.0 & $17.3 -37.1$ & $6.2\pm2.9$ & \\
$r$
& $0.38 \pm 0.08$ & 1.3 & $3.0 - 4.9$ & $0.37\pm 0.10$
& $0.39^{+0.15}_{-0.13}$\\
\end{tabular}

{\small $^a$ $B^0\to K^{*0}\gamma$

$^b$ $B^+\to K^{*+}\gamma$}
\end{table}

\begin{figure}
\unitlength=0.9mm
\begin{picture}(150,150)
\put(10,100){\line(1,0){50}}
\put(10,120){\line(1,0){50}}
\put(35,120){\circle*{5}}
\multiput(32.5,130)(0,-10){2}{\begin{picture}(5,10)
\put(2.5,10){\oval(5,5)[r]}
\put(2.5,5){\oval(5,5)[l]}\end{picture}}
\put(5,120){\large$b$}
\put(5,100){\large$\bar q$}
\put(5,110){\large$B$}
\put(65,120){\large$s$}
\put(65,100){\large$\bar q$}
\put(65,110){\large$K_2^*$}
\put(43,140){\large$\gamma$}
\put(0,85){\small FIG. 1. Lowest order vertex function $\Gamma^{(1)}$
corresponding to Eq.~(\ref{gam1}). }
\put(10,20){\line(1,0){50}}
\put(10,40){\line(1,0){50}}
\put(25,40){\circle*{5}}
\put(25,40){\thicklines \line(1,0){20}}
\multiput(25,40.5)(0,-0.1){10}{\thicklines \line(1,0){20}}
\put(25,39.5){\thicklines \line(1,0){20}}
\put(45,40){\circle*{1}}
\put(45,20){\circle*{1}}
\multiput(45,40)(0,-4){5}{\line(0,-1){2}}
\multiput(22.5,50)(0,-10){2}{\begin{picture}(5,10)
\put(2.5,10){\oval(5,5)[r]}
\put(2.5,5){\oval(5,5)[l]}\end{picture}}
\put(5,40){\large$b$}
\put(5,20){\large$\bar q$}
\put(5,30){\large$B$}
\put(65,40){\large$s$}
\put(65,20){\large$\bar q$}
\put(65,30){\large$K_2^*$}
\put(33,60){\large$\gamma$}
\put(90,20){\line(1,0){50}}
\put(90,40){\line(1,0){50}}
\put(125,40){\circle*{5}}
\put(105,40){\thicklines \line(1,0){20}}
\multiput(105,40.5)(0,-0.1){10}{\thicklines \line(1,0){20}}
\put(105,39,5){\thicklines \line(1,0){20}}
\put(105,40){\circle*{1}}
\put(105,20){\circle*{1}}
\multiput(105,40)(0,-4){5}{\line(0,-1){2}}
\multiput(122.5,50)(0,-10){2}{\begin{picture}(5,10)
\put(2.5,10){\oval(5,5)[r]}
\put(2.5,5){\oval(5,5)[l]}\end{picture}}
\put(85,40){\large$b$}
\put(85,20){\large$\bar q$}
\put(85,30){\large$B$}
\put(145,40){\large$s$}
\put(145,20){\large$\bar q$}
\put(145,30){\large$K_2^*$}
\put(133,60){\large$\gamma$}
\put(0,5){\makebox[14cm][s]{\small FIG. 2. Vertex function $\Gamma^{(2)}$
corresponding to Eq.~(\ref{gam2}). Dashed lines represent the}}
\put(0,0) {\makebox[14cm][s]{\small  effective   potential ${\cal V}$ in
Eq.~(\ref{qpot}). Bold lines denote the negative-energy part of the} }
\put(0,-5){\small quark propagator. }

\end{picture}
\end{figure}

\end{document}